\documentclass[aip,pre,onecolumn,superscriptaddress,floatfix,12pt]{revtex4}
\usepackage{graphicx}
\usepackage{amsmath}
\usepackage{color}
\begin{document}

\title{Sequence transferable coarse-grained model of amphiphilic copolymer}

\author{Chathuranga C. De Silva}
\affiliation{Department of Chemical Engineering, Columbia University, New York USA}
\author{Porakrit Leophairatana}
\affiliation{Department of Chemical Engineering, Columbia University, New York USA}
\author{Takahiro Ohkuma}
\affiliation{Central Research Division, Bridgestone Corporation, Kodaira, Tokyo 187-8531 Japan}
\affiliation{Max-Planck Institut f\"ur Polymerforschung, Ackermannweg 10, 55128 Mainz Germany}
\author{Jeffrey T. Koberstein}
\affiliation{Department of Chemical Engineering, Columbia University, New York USA}
\author{Kurt Kremer}
\affiliation{Max-Planck Institut f\"ur Polymerforschung, Ackermannweg 10, 55128 Mainz Germany}
\author{Debashish Mukherji}
\affiliation{Max-Planck Institut f\"ur Polymerforschung, Ackermannweg 10, 55128 Mainz Germany}

\date{\today}

\begin{abstract}
Polymer properties are inherently multi-scale in nature, where delicate local interaction
details play a key role in describing their global conformational behavior.
In this context, deriving coarse-grained (CG) multi-scale models for polymeric liquids is a non-trivial task.
Further complexities arise when dealing with copolymer systems with varying microscopic
sequences, especially when they are of an amphiphilic nature. In this work, we derive
a segment-based generic CG model for amphiphilic copolymers consisting of repeat units of hydrophobic (methylene)
and hydrophilic (ethylene oxide) monomers. The system is a simulation analogue of polyacetal
copolymers [Samanta et al., Macromolecules 49, 1858 (2016)].
The CG model is found to be transferable over a wide range of copolymer sequences and also to be consistent with
existing experimental data.
\end{abstract}

\maketitle

\section{Introduction}
\label{sec:intro}

Polymer processing requires a detailed understanding of their structure-property relationships, 
which not only is fundamentally challenging but also has a wide range of applications ranging from 
physics to biology \cite{cohen10natmat,mukherji14natcom}. In this context, smart responsive polymers 
serve as excellent candidates. A polymer is commonly known as ``smart responsive" when a small change in
external stimulus can drastically change its structure, function and/or stability. Furthermore, in these systems, 
where the relevant energy scale is of the order of the thermal energy $k_{\rm B}T$, large conformational and local compositional fluctuations 
play a delicate role dictating their properties \cite{peter1,noid}. When the external stimulus is temperature $T$, these polymers are 
referred to as thermoresponsive polymers \cite{hoogenboom,koberstein16mac}. 
One of the interesting classes of thermoresponsive polymers is those that remain expanded at
low $T$ and collapse into compact objects when $T > T_\ell$, where $T_\ell$ is
referred to as a lower critical solution temperature (LCST). The properties of LCST polymers are usually dictated by hydrogen 
bonding of water molecules with polymer segments, which break down when $T> T_\ell$. During this process, 
when hydrogen hydrogen bonds are broken, water molecules are expelled from near the polymer structure and thus 
gain a large amount of translational entropy. Therefore, simply speaking the polymer chain will collapse whenever the total translational 
entropy gained by water molecules is larger than the conformational entropy lost by the polymer upon collapse \cite{degennesbook}.

For a given homopolymer structure there exists a characteristic $T_\ell$ \cite{WuMac98,tucker12mac,RichteringJPS13}.
This $T_\ell$ can, however, be tuned by introducing more hydrophobic or more hydrophilic units along the homopolymer 
backbone \cite{koberstein16mac,COPOLexp00,COPOLexp06,rad15,joejcp,tiago17jcp}. 
In particular, $T_\ell$ increases (decreases) with increasing hydrophilic (hydrophobic) 
units. Moreover, the changes in $T_\ell$ are usually difficult to predict and non-linear with changing copolymer 
sequences \cite{COPOLexp00,COPOLexp06,rad15}. However, in a recent work, based on the acetal linkage of methylene and ethylene oxide units, 
it has been shown that $T_\ell$ can be linearly tuned by adjusting the fractions of methylene and ethylene oxide monomers 
along the polymer backbone \cite{koberstein16mac}. 
In addition to predictability with changes in sequence, the acetal-based copolymers are 
advantageous due to their biodegradable properties. Moreover, these polymers 
show strong chain length effects dependent on end-group nature (see Fig. 5 in Ref. [\onlinecite{koberstein16mac}]). 
For example, it has been shown that a molecular weight $M_{\rm w} \sim 10^4$ g/mol is necessary to avoid chain length effects, 
which corresponds to an end-to-end distance $R_{\rm ee} \sim 10$ nm for a poly(ethylene oxide). Within a simulation setup this would 
require a box size of at least 15 to 20 nm. For a water box, with number density of 32 water molecules per cubic nm, more than $10^5$ 
water molecules would be needed (or equivalent of $\sim 3 \times 10^5$ atoms). 
This poses a serious challenge for all-atom simulations of these systems and motivates the development of accurate, lower resolution 
CG model for further theoretical investigations. While there are studies in deriving implicit and explicit solvent CG models 
of PEO \cite{peo1,peo2,peo3}, in this work we derive a sequence transferable, systematic CG model of 
an analogue of polyacetal that represents each copolymer units with single CG beads. We employ the model to study different
copolymer architectures that are complementary to the earlier experimental results \cite{koberstein16mac}. Additionally, we also 
propose a broad range of molecular morphologies, going beyond available experimental data.
Note that here we discuss a case where hydrophobicity (or hydrophilicity) is tuned along the backbone and not added as a side group. 
However, there are cases, such as poly(N-isopropylmethacrylamide) where an extra methylene group added as a side group to 
poly(N-isopropylacrylamide) increases $T_{\ell}$ \cite{pnipmam}. The possible reason behind an added hydrophobic group increases $T_{\ell}$ can be 
attributed to the solvent structure around the chain \cite{tiago17jcp}. 

\section{Model and method}
\label{sec:model}

\begin{figure}[ptb]
\includegraphics[width=0.49\textwidth,angle=0]{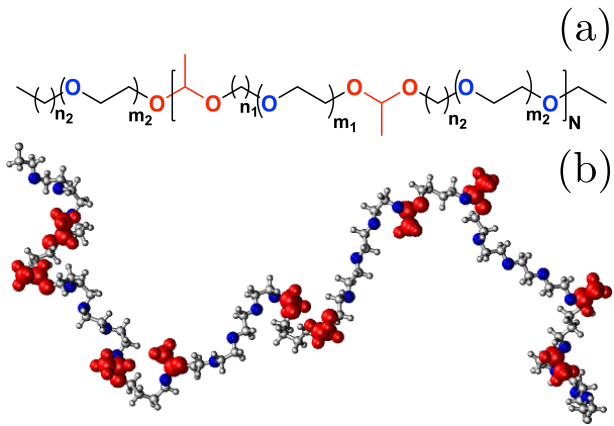}
\caption{Part (a) shows a chemical structure of the polyacetal system synthesized in an earlier experiment terminated with CH$_3$ group \cite{koberstein16mac}. 
The hydrophobic methylene units ($n_1$ and $n_2$) and hydrophilic ethylene oxide units ($m_1$ and $m_2$) 
are tuned to obtain different amphiphilic sequences. The monomer is represented within the large bracket, 
where $N$ is the number of repeat units. Part (b) shows a simulation snapshot of a chain with $N=5$ in water at 290 K 
and for $n_1=4$, $m_1=0$, $n_2 =2$, and $m_2=3$.
\label{fig:polyace}}
\end{figure}

We start by commenting on the underlying reference all-atom simulations. 
For this purpose, the chemical structures are taken to be the same as in experiment \cite{koberstein16mac}.
In Fig.~\ref{fig:polyace}(a) we show a typical chemical configuration of the polyacetal copolymer architecture.
In this amphiphilic structure, $T_\ell$ is tuned by changing the numbers of methylene ($n_1$ and $n_2$) and ethylene oxide 
($m_1$ and $m_2$) units. For computational simplicity of the computing, we have chosen the repeat unit of the chain as $N = 5$ (see Fig.~\ref{fig:polyace}(b)).
%The standard OPLS force field, consisting of explicit atoms, is used to model the polymer structure \cite{opls}, which is solvated in a 
%water box consisting of 17000 water molecules described by the SPC/E water model \cite{spce}. 
Here we consider a copolymer structure represented with $n_1=4$, $m_1=0$, $n_2 =2$, and $m_2=3$.
Note that the molecular weight of the simulated chain presented in Fig.~\ref{fig:polyace}(b) is almost one order of magnitude 
smaller than the $M_{\rm w}$ needed to avoid strong chain length effects \cite{koberstein16mac}. 
However, it is noteworthy that the system size effects in the experiments are associated with the end group effects. 
Therefore, to avoid the system size effect and to make a reasonable estimate of $T_\ell$, we have terminated the ends of copolymer with an inert CH$_3$ groups, 
as shown in Fig. \ref{fig:polyace}(b). As will be discussed in the latter part of this manuscript, this approximation allows us 
to obtain a reasonable $T_\ell$, while not attempting to make any claims on the system size effects in simulations. Moreover, 
this small chain length can not be used for any structural predictions that only happen for the longer chains.

All-atom molecular dynamics (MD) simulations are performed using the GROMACS package \cite{gro}.
The temperature is varied from 290 K to 320 K using velocity rescaling with a coupling constant 0.5 ps \cite{v-rescale}.
Each of these simulations are performed for 50 ns production runs, which is at least one order of magnitude larger than the longest 
relaxation time. The average is taken over the last 10 ns of MD data. The electrostatics are treated using Particle Mesh Ewald \cite{pme}. 
The interaction cut-off for non-bonded interactions is chosen as 1.0 nm.
The simulations are performed with a constant pressure ensemble, where the
the pressure is controlled using a Berendsen barostat \cite{berend} with a coupling time of 0.5 ps and 1 atm pressure.
The time step for the simulations is set to 2 fs and the equations of motion are integrated using the leap-frog algorithm.
LINCS algorithm is used to constraint all bond vibrations \cite{lincs}. 

We find the gyration radius $R_{\rm g}$ to be $1.33\pm0.15$ nm (for $T=290$ K), $1.38\pm0.16$ nm (for $T=300$ K), $0.97\pm0.04$ nm (for $T=310$ K), 
and $0.98\pm0.09$ nm (for $T=320$ K) for the system presented in Fig. \ref{fig:polyace}(b). 
There is a reasonably sharp change in $R_{\rm g}$ between 300K and 310 K, suggesting $T_\ell$ to be between 300 K and 310 K.
The experimental phase diagram, however, suggests $T_\ell\sim 320$ K for the same copolymer sequence. 
Therefore, our all-atom data underestimates $T_\ell$ by about 10$-$20 K. 

Non-bonded CG potentials are derived using the structure-based techniques for solutions \cite{originibi1,ibi,cibi}. 
To derive the CG model, which is sequence transferable, we make two assumptions: 
1) Upon changing composition and sequencing along the backbone, there are no cross-correlations between 
different monomer units. In this context, experiments have shown that $T_\ell$ changes linearly with changing fractions 
of hydrophobic or hydrophilic units  \cite{koberstein16mac}. 
Therefore, presenting a situation where a segment based coarse-graining (only building CG models 
based on separate simulations of different monomer units) can be performed, which can later be incorporated into a polymer 
chain with different sequences. 
\begin{figure}[ptb]
\includegraphics[width=0.49\textwidth,angle=0]{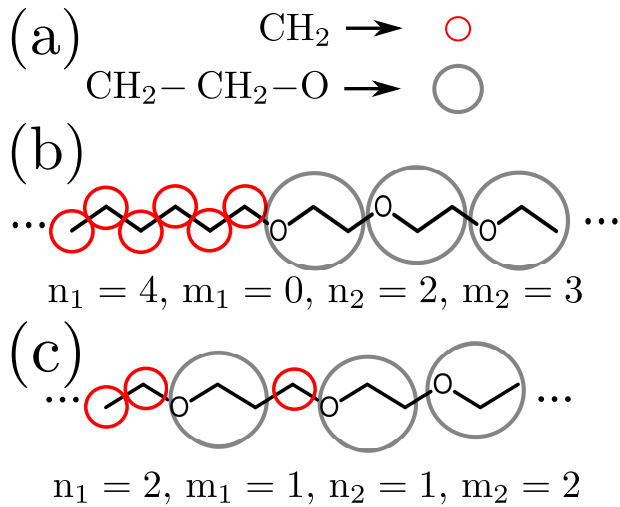}
\caption{Part (a) shows a mapping scheme of methylene and ethylene oxide monomers. Part (b) and (c) show 
monomer segments of two different copolymer sequences, namely $n_1=4$, $m_1=0$, $n_2 =2$, and $m_2=3$ and 
$n_1=2$, $m_1=1$, $n_2 =1$, and $m_2=2$. 
\label{fig:cging}}
\end{figure}
In this context, a generic study of copolymer sequences, 
within a mean-field picture, have shown linear or non-linear pair interpolation depending of the interaction parameters \cite{joejcp}.
2) The acetal linker (represented by red in Fig.~\ref{fig:polyace}) only contributes to a
negligible shift in $T_\ell$. Therefore, we do not incorporate an acetal linker in our simulations. We note that 
one might use a hard sphere type representation of the acetal unit, although this
is beyond the motivation of the present study. The step-by-step procedure for building the CG model is:
For the non-bonded interaction, we perform all-atom simulation of different monomer unitss solvated in water, 
with the mapping scheme presented in Fig. \ref{fig:cging}(a). 
\begin{figure}[ptb]
\includegraphics[width=0.49\textwidth,angle=0]{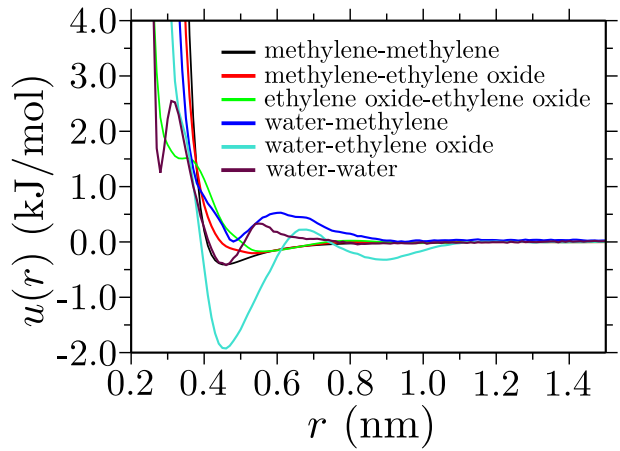}
\caption{Coarse-grained interaction potential $u(r)$ between different pairs of solution components.
Results are obtained for $T=320$ K, which corresponds to a collapsed structure for a 
sequence with $n_1=4$, $m_1=0$, $n_2 =2$, and $m_2=3$.
\label{fig:vr}}
\end{figure}

\section{Results and discussions}
\label{sec:res}

The resultant interactions potentials \cite{potential}, shown in Fig. \ref{fig:vr}, are used to simulate the whole range of 
copolymer sequences. Note that the derived potentials from the monomer level is used in simulations of polymer chain
with one-four exclusion of bonded neighbors. Importantly, we derive only one set of $u(r)$ that we will
use for a wide range of copolymer sequences. Bonded interactions are obtained by Boltzmann inversion of bonded, angle and 
dihedral distributions. CG simulations are also performed in GROMACS package \cite{gro}, where the 
interaction cut-off is chosen as 1.8 nm and the time step is chosen as 2 fs.
Simulations are performed for a 100 ns long CG MD trajectory. 

\begin{table}[ptb]
\centering
\caption{Copolymer conformations with different sequences (as shown in the first four columns). 
In the fifth column we present methylene mole fraction $x_m$ along the copolymer backbone,
which is defined as $x_m = (n_1+n_2)/(n_1+m_1+n_2+m_2)$. The data is shown for $T = 320$ K. 
The transition temperature in our reference all-atom simulations is 
about 20 K lower than the experimental transition temperature. Therefore, we present the 
extrapolated polymer conformations from experiments, with the transition temperature shifted lower by 20 K.
In the last column, we present corresponding polymer conformation from CG simulations.
Note that the polymer conformations are identified by calculating their 
single chain form factor $S(q)$, which shows $S(q)\sim q^{-5/3}$ for an expanded chain and $S(q)\sim q^{-4}$
for a collapsed globule.}
\label{tab:exp_sim}
\begin{tabular}{|l|l|l|l|l|l|l|}\hline
$n_1$ & $m_1$ & $n_2$ & $m_2$ & $~~~x_m~~~$ & Experiment \cite{koberstein16mac}  &  Simulation \\ \hline
\hline                                                                                                    
  4   &  0    &  2    &   0           &  1.00    & Turbid                                 & Globule \\ \hline        
  4   &  0    &  2    &   1           &  0.86    & Turbid                                 & Globule \\ \hline       
  4   &  0    &  2    &   2           &  0.75    & Turbid                                 & Globule \\ \hline       
  4   &  0    &  2    &   3           &  0.67    & Turbid                                 & Globule \\ \hline       
  4   &  0    &  2    &   4           &  0.60    & Turbid                                 & Globule \\ \hline       
  4   &  0    &  2    &   5           &  0.54    & Turbid                                 & Globule \\ \hline       
\hline                                                                                                    
  2   &  1    &  2    &   1           &  0.67    & Turbid                                 & Globule  \\ \hline      
  2   &  1    &  2    &   2           &  0.57    & Turbid                                 & Globule  \\ \hline      
  2   &  1    &  2    &   3           &  0.50    & Clear                                  & Lamellar \\ \hline     
  2   &  1    &  2    &   4           &  0.44    & Clear                                  & Expanded     \\ \hline  
  2   &  1    &  2    &   5           &  0.40    & Clear                                  & Expanded     \\ \hline  
\hline                                                                                                    
  2   &  2    &  2    &   1           &  0.57    & Clear                                  & Globule \\ \hline      
  2   &  2    &  2    &   2           &  0.50    & Clear                                  & Expanded \\ \hline      
  2   &  2    &  2    &   3           &  0.44    & Clear                                  & Expanded \\ \hline      
  2   &  2    &  2    &   4           &  0.40    & Clear                                  & Expanded \\ \hline      
  2   &  2    &  2    &   5           &  0.36    & Clear                                  & Expanded \\ \hline      
\end{tabular}                                                                                                       
\end{table}

\begin{figure}[ptb]
\includegraphics[width=0.49\textwidth,angle=0]{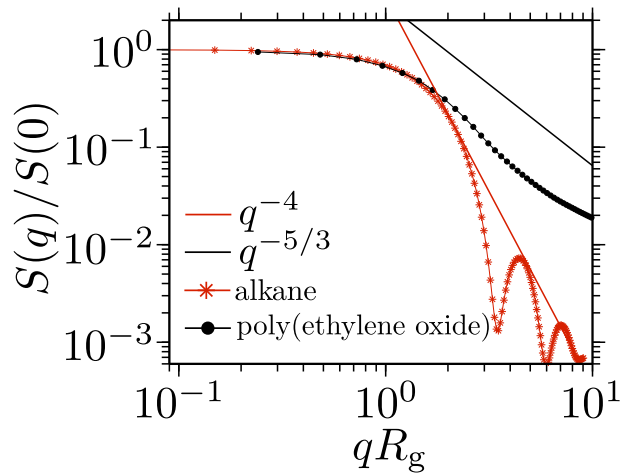}
\caption{Single chain static structure factor $S(q)$ for an alkane chain and a poly(ethylene oxide) chain in water. 
The power law $q^{-5/3}$ represents a good solvent chain and $q^{-4}$ shows a well collapsed spherical globule.
\label{fig:s_of_q}}
\end{figure}

We start by investigating the conformation of 
alkane and poly(ethylene oxide) chains in water. For this purpose, the 
molecular weight is chosen as $M_{\rm w} = 10^4$ g/mol.
The CG model reproduces the expected conformations for both these chains, i.e., an expanded chain for poly(ethylene oxide) 
with $R_{\rm g} = 3.91 \pm 0.46$ nm, which is consistent with the previous simulations \cite{martini} 
and a collapsed structure for the alkane chain with $R_{\rm g} = 1.18\pm0.05$ nm.
The structures are characterized by their single chain structure 
factor $S(q)$ shown in Fig.~\ref{fig:s_of_q}. For example, expanded chain shows a scaling 
law $q^{-1/\nu}$ with $\nu = 3/5$ being the Flory's exponent for a good solvent chain (see black curve in Fig.~\ref{fig:s_of_q}). 
When a polymer collapses, its conformation is well described by a hard sphere scattering function with a scaling 
law $q^{-4}$ (see red curve in Fig.~\ref{fig:s_of_q}) \cite{degennesbook}. 

Now we focus on the main theme of this work, i.e., to study the conformational behavior of complex amphiphilic structures 
and their comparison to known experimental data obtained from the polyacetal system \cite{koberstein16mac}. 
In Table~\ref{tab:exp_sim} we summarize data from experimental copolymers and its comparison to these CG simulations. 
Note that turbid solutions found by experiment indicate collapsed structures, while clear solutions are associated with expanded conformations.
Comparing experimental and simulation data in the last two columns, we find that the polymer conformation is reasonably well reproduced by the CG model,
except for two cases. We would also like to point out that for methylene mole fraction $x_m > 50\%$, chains are always collapsed as 
shown in the column five of Table~\ref{tab:exp_sim}.

\begin{figure*}[ptb]
\includegraphics[width=0.70\textwidth,angle=0]{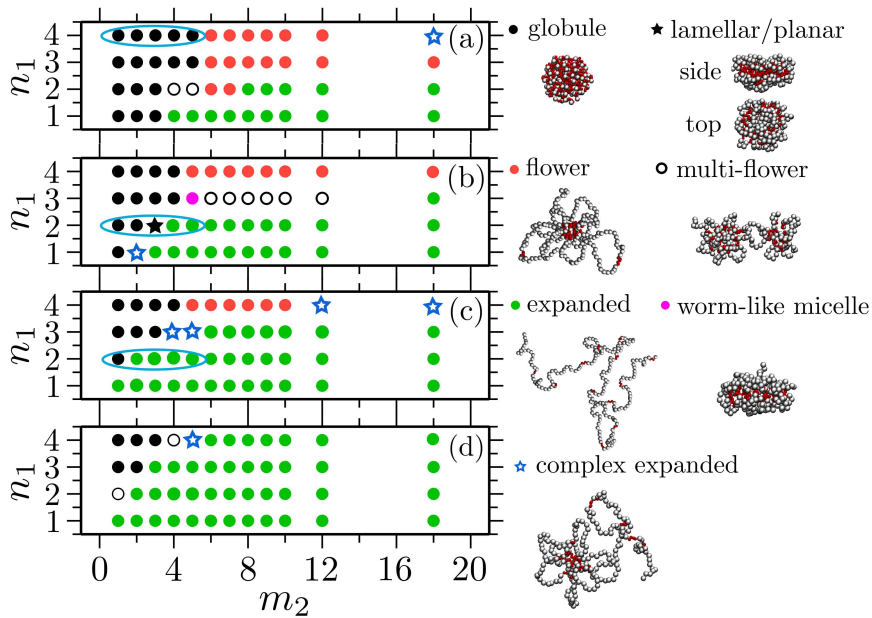}
\caption{A representative phase diagram of the amphiphilic copolymers with different sequences. The data is shown for 
(a) $m_1$ = 0, (b) $m_1$ = 1, (c) $m_1$ = 2, and (d) $m_1$ = 3 with $n_2 =$ 2 and varying $m_2$ and $n_1$. 
Every symbol in these figures represent one configuration with the color code consistent with the configurations presented in the caption. 
Ethylene oxide beads are rendered in silver and methylene units are represented by red spheres. Highlighted oval configurations correspond to
the experimental copolymers shown in Table~\ref{tab:exp_sim}.
\label{fig:phase_diag}}
\end{figure*}

We also would like to draw attention to the fact that it is generally challenging to incorporate composition (or sequence) and 
temperature transferability in a CG model. Table \ref{tab:exp_sim} demonstrates that our CG model is well transferable over a wide range of 
copolymer sequences. Further testing reveals that the CG model only reproduces reliable structures for this particular thermodynamic state 
of parameterization, i.e., 320 K. This is not surprising given that the many-body potential of mean force (PMF) is state point-dependent. 
However, there are examples where the many-body features are weak and thus the CG model can be $T$ transferable. Moreover, hydrogen 
bonding nature of the interactions, as in the case of polyacetal, adds up an additional complexity, making the many-body effect more relevant and thus CG potential 
become less transferable. 

The advantage of a segment-transferable CG model, which originate because of the lack in cross-correlation between different monomer units 
over large length scales along the polymer backbone, is not only that it allows a rather consistent comparison with the experiments \cite{koberstein16mac}, 
but that is also enables structural predictions to be made for several more macromolecular architectures.
Therefore, the CG model presented here can be viewed as a molecular toolbox to investigate the properties of different 
polymer architectures for advanced functional uses \cite{potential}. In this context, it has been previously predicted that
amphiphilic copolymers can exhibit interesting structures \cite{halperin,halperin2}, which was later studied by generic 
simulations \cite{flower1} and Monte Carlo simulations \cite{flower2}. These complex structures are highly interesting 
for biomedical applications, such as drug delivery and materials for tissue engineering \cite{cohen10natmat,flower2}.
Therefore, to investigate a broad range of conformational properties of polyacetal-based systems, we have constructed a 
set of 192 copolymer configurations in water with varying amphiphilic sequences. Each of these simulations were performed 
for 100 ns in CG units, generating a total of about 15 $\mu$s of CG MD data. 

In Fig.~\ref{fig:phase_diag} we present three sets of representative phase diagrams for the copolymer 
representing with varying $n_1$, $n_2$, $m_1$, and $m_2$. It can be seen that depending on the sequences, we observe a variety of copolymer configurations. 
For example, we find flower, micellar, multiple-flower, lamellar, and/or worm-like micellar structures. This presents 
a fully flexible and versatile molecular tool box for the simulation of these amphiphilic systems. 
In addition to the 192 configurations presented here many additional configurations may be generated 
by employing distinct combinations of hydrophobic and hydrophilic units. Furthermore, the applicability of this CG model 
is not only restricted to the linear amphiphilic chains, but may also be useful to study the solvation and/or 
aggregation of branched or brush like copolymers \cite{weilAML}. 

\section{Conclusions}
\label{sec:conc}

We have presented a protocol to obtain a sequence-transferable coarse-grained (CG) model for amphiphilic copolymer architectures.
Our CG model was derived from a segment (monomer) based level and then was translated into longer chain simulations with
different amphiphilic sequences. The derived CG model was validated by one-to-one comparison with experimental data \cite{koberstein16mac}.
Additionally, we also make several structural predictions that go beyond the polymers synthesized in experiments. 
While the derived CG model is sequence transferable, it shows poor temperature transferability.
This is due to the hydrogen bonded nature of the underlying interaction details, which leads to a complex 
many-body effects that can not be captured within structure-based CG model. Here, however, parameterization of 
a CG model at different temperatures $T$ can be performed to obtain a $T$ dependent interaction capturing correct 
nature of hydrogen bonded interaction \cite{stevens,joe}. Furthermore, the segment transferability is important because one 
set of CG potentials can describe, predict, and validate many amphiphilic polymer architectures. Therefore, the potentials presented here can be viewed as a molecular toolbox for 
a wide range of alkane and ethylene oxide based architectures. 

\section{Acknowledgement}
\label{sec:ack}

We thank Burkhard D\"unweg and Carlos M. Marques for countless interesting discussions and 
Tiago E. de Oliveira for the help with the $\mathcal{C}-$IBI scripts implemented in the VOTCA 
package \cite{votca} and all atom force field. We further acknowledge Joseph Rudzinski, Nancy Carolina Forero-Martinez, and Torsten St\"uhn for critical reading of
the manuscript. C.D.S., P.L. and J.T.K. wish to acknowledge support by the National Science Foundation under grant DMR-1206191. 
D.M. thank Joachim Dzubiella for discussions related to his manuscript draft on CG model of
poly(ethyle oxide) homopolymer in water \cite{joe}.


\begin{thebibliography}{25}

\bibitem{cohen10natmat}
M. A. Cohen-Stuart, W. T. S. Huck, J. Genzer, M. M\"uller, C. Ober, M. Stamm, G. B. Sukhorukov,
I. Szleifer, V. V. Tsukruk, M. Urban, F. Winnik, S. Zauscher, I. Luzinov, and S. Minko,
Nature Materials {\bf 9}, 101 (2010).

\bibitem{mukherji14natcom}
D. Mukherji, C. M. Marques, and K. Kremer,
Nat. Commun. {\bf 5}, 4882 (2014).

\bibitem{peter1}
C. Peter and K. Kremer,
Soft Matter {\bf 5}, 4357 (2009).

\bibitem{noid}
W. G. Noid,
J. Chem. Phys. {\bf 139}, 090901 (2013).

\bibitem{hoogenboom}
Q. Zhang and R. Hoogenboom,
Prog. Pol. Science {\bf 48}, 122 (2015).

\bibitem{koberstein16mac}
S. Samanta, D. R. Bogdanowicz, H. H. Lu, and J. T. Koberstein,
Macromolecules {\bf 49}, 1858 (2016).

\bibitem{degennesbook}
P. G. de Gennes,
{\it Scaling Concepts in Polymer Physics}
(Cornell University Press, London, 1979).

\bibitem{WuMac98}
X. Wang, X. Qiu, and C. Wu,
Macromolecules {\bf 31}, 2972 (1998).

\bibitem{tucker12mac}
A. K. Tucker, and M. J. Stevens,
Macromolecules {\bf 45}, 6697 (2012).

\bibitem{RichteringJPS13}
H. Kojima, F. Tanaka, C. Scherzinger, and W. Richtering,
J. Polym. Sci., Part B: Polym. Phys. {\bf 51}, 1100 (2013).

\bibitem{COPOLexp00}
A. S. Hoffman, P. S. Stayton, V. Bulmus, G. Chen, J. Chen, C. Cheung, A. Chilkoti, Z. Ding, L. Dong, R. Fong, C. A. Lackey, C. J. Long, M. Miura, J. E. Morris, N. Murthy, 
Y. Nabeshima, T. G. Park, O. W. Press, T. Shimoboji, S. Shoemaker, H. J. Yang, N. Monji, R. C. Nowinski, C. A. Cole, J. H. Priest, J. M. Harris, K. Nakamae, T. Nishino, and T. Miyata,
J. Biomed. Mat. Res. {\bf 52}, 577 (2000).

\bibitem{COPOLexp06}
Z. Shen, K. Terao, Y. Maki, T. Dobashi, G. Ma, and T. Yamamoto,
Colloid Polym. Sci. {\bf 284}, 1001 (2006).

\bibitem{rad15}
M. Radecky, J. Spevacek, A. Zhigunov, Z. Sedlokova, and L. Hanykova,
Eur. Polym. J. {\bf 68}, 68 (2015).

\bibitem{joejcp}
B. Schulz, R. Chudoba, J. Heyda, and J. Dzubiella,
J. Chem. Phys. {\bf 143}, 243119 (2015).

\bibitem{tiago17jcp}
T. E. de Oliveira, D. Mukherji, K. Kremer, and P. A. Netz,
J. Chem. Phys. {\bf 146}, 034904 (2017).

%\bibitem{jeppesen95epl}
%C. Jeppesen and K. Kremer,
%Eur. Phys. Lett. {\bf 34}, 563 (1996).

\bibitem{peo1}
E. Choi, J. Mondal, and A. Yethiraj
J. Phys. Chem. B {\bf 118}, 323 (2014).

\bibitem{peo2}
S. Nawaz and P. Carbone
J. Phys. Chem. B {\bf 118}, 1648 (2014).

\bibitem{peo3}
S. Wang and R. G. Larson
Macromolecules {\bf 48}, 7709 2015.

\bibitem{pnipmam}
J. Gernandt, G. Frenning, W. Richtering and P. Hansson,
Soft Matter {\bf 7}, 10327 (2011).

\bibitem{gro}
S. Pronk, S. Pall, R. Schulz, P. Larsson, P. Bjelkmar, R. Apostolov, M. R. Shirts, J. C. Smith, P. M. Kasson, D. Van Der Spoel, B. Hess, and E. Lindahl,
Bioinformatics {\bf 29}, 845 (2013).

\bibitem{v-rescale}
G. Bussi, D. Donadio, and M. Parrinello,
J. Chem. Phys. {\bf 126}, 014101 (2007).

\bibitem{pme}
U. Essmann, L. Perera, M. L. Berkowitz, T. Darden, H. Lee, and L. G. A. Pedersen,
J. Chem. Phys. {\bf 103}, 8577 (1995). 

\bibitem{berend}
H. J. C. Berendsen, J. P. M. Postma, W. F. van Gunsteren, A. DiNola, and J. R. Haak,
J. Chem. Phys. {\bf 81}, 3684 (1984).

\bibitem{lincs}
B. Hess, H. Bekker, H. J. C. Berendsen, and J. G. E. M. Fraaije,
J. Comput. Chem. {\bf 18}, 1463 (1997).

\bibitem{originibi1}
W. Tsch\"op, K. Kremer, J. Batoulis, T. B\"urger, and O. Hahn,
Acta Polymer {\bf 49}, 61 (1998); Acta Polymer {\bf 49}, 75 (1998).

\bibitem{ibi}
D. Reith, M. P\"utz, and F. M\"uller-Plathe,
J. Comput. Chem. {\bf 24}, 1624 (2003).

\bibitem{cibi}
T. E. de Oliveira, P. A. Netz, K. Kremer, C. Junghans, and D. Mukherji,
J. Chem. Phys. {\bf 144}, 174106 (2016).

\bibitem{potential}
Interaction potentials and all the automated scripts to generate different molecular architectures 
will be compiled in a web domain located at the MPIP.

\bibitem{martini}
H. Lee, A. H. de Vries, S.-J. Marrink and R. W. Pastor,
J. Phys. Chem. B {\bf 113}, 13186 (2009).

\bibitem{halperin}
A. Halperin,
Macromolecules {\bf 24}, 1418 (1991).

\bibitem{halperin2}
O. V. Borisov and A. Halperin,
Macromolecules {\bf 29}, 2612 (1996).

\bibitem{flower1}
J. Zhang, Z.-Y. Lu and Z.-Y. Sun
Soft Matter {\bf 9}, 1947 (2013).

\bibitem{flower2}
V. Hugouvieux, M. A. V. Axelos, and M. Kolb
Macromolecules {\bf 42}, 392 (2009).

\bibitem{weilAML}
A. Riegger, C. Chen, O. Zirafi, N. Daiss, D. Mukherji, K. Walter, Y. Tokura, B. Stoeckle, K. Kremer, F. Kirchhoff, D. Y. W. Ng, P. C. Hermann, J. Münch, and T. Weil,
ACS Macro Lett. {\bf 6}, 241 (2017).

\bibitem{stevens}
L. J. Abbott and M. J. Stevens,
J. Chem. Phys. {\bf 143}, 244901 (2016).

\bibitem{votca}
S. Y. Mashayak, M. N. Jochum, K. Koschke, N. R. Aluru, V. R\"uhle, and C. Junghans,
PLoS one {\bf 10}, e131754 (2015).

\bibitem{joe}
J. Dzubiella et al., {\it private communications} (2017).

\end{thebibliography}
\end{document}